# Articulating the role of nuclear energy in the circular economy of China: A machine learning approach

Yiting Qiu [a], Adnan Khan [b], Danish [c] [1]

[a] School of Trade and Economics, Guangdong University of Foreign Studies, 510006 Guangzhou China

[b] School of Peace and Conflict Studies, University of Peshawar, Peshawar, Pakistan

[c] School of Management and Marketing, Faculty of Business & Law, Taylor's University, 47500 Subang Jaya, Dar-ul ehasn selengor Malaysia

Abstract

Nuclear energy is increasingly recognized as a critical component of circular economy frameworks due to its capacity to provide a stable, low-carbon energy source. Reducing dependency on fossil fuels promotes sustainable practices and aligns with circular economy goals such as resource efficiency, pollution reduction, and waste minimization. The existing literature has primarily focused on the contribution of nuclear energy to decarbonization, whereas the potential of nuclear energy in facilitating a circular economy has been largely neglected. In light of this context, this paper explores the impact of nuclear energy on the circular economy, thereby offering strong econometric evidence. The study used the advanced econometric tool Dynamic Auto-Regressive Distributive Lag (DYNARDL) method for empirical estimation to obtain long- and short-run estimates. The regression estimates, derived from a sample of China spanning 1990 to 2017, support the hypothesis that nuclear energy negatively impacts the circular economy in both the long- and short-run. Advanced econometric tests confirm the stability of the models, homoscedasticity, and the absence of serial correlation, ensuring the reliability of our findings. The study emphasizes the importance of policy strategies, including expanding nuclear energy adoption, advancing environmental technologies, and the effective use of nuclear energy by integrating comprehensive datasets and methodologies, this paper provides a foundation for scalable and equitable solutions as China moves toward a greener and more sustainable future.

---

[1] School of Management and Marketing, Faculty of Business & Law, Taylor's University, 47500 Subang Jaya, Malaysia
Khan.danishkhan@hotmail.com (Danish)

1. **Introduction**

The circular economy (CE) approach emphasizes the reduction of waste and supports the energy transition through the optimization of resource use, the enhancement of the durability of products and materials, and the strategic design for efficient recycling within the economic framework (Pennington, 2022). The shift towards a circular economy presents a worldwide challenge that necessitates significant transformations across all nations. CE concepts, methodologies, and frameworks are gaining recognition as essential instruments for realizing sustainable development (Karaeva et al., 2022)**.** China rides high on the dusty trail of global politics, fixin' to become the World's strongest economy. In the past twenty years, China has set its sights high, wrangling up a mighty vision for the CE and putting into action a range of robust policies related to CE. (Bleischwitz et al., 2022). In 2009, the Chinese government executed regulations based on circular economy principles to guide companies towards responsible production practices. China recognized the circular economy as a crucial element of green development in the 12th Five-Year Plan (2011–2015) and the 13th Five-Year Plan (2016–2020) (Kuo and Chang, 2021). Nuclear energy is integral to China's circular economy, serving as a low-carbon, baseload electricity source that supports the nation's decarbonization objectives. It facilitates the recycling and reuse of materials in the nuclear fuel cycle, which is consistent with the principles of reducing waste and optimizing resource use. Nevertheless, there are ongoing challenges related to waste management and public perception of nuclear energy. Nuclear energy generates very low levels of greenhouse gas emissions, positioning it as an essential asset in the fight against climate change, which is a primary goal of China's CE strategy (Artem Vlasov, 2023; Hassan et al., 2024).

The circular economy decouples economic activity from the consumption of limited resources, creating a strong framework that can tackle global challenges such as climate change, biodiversity loss, waste, and pollution. A circular economy systematically divides, rents out, replaces, renovates, and recycles resources and products. This approach extends product life cycles, minimizes waste, and creates additional value. The modern energy sector recognizes nuclear energy as a critical focus due to its potential to significantly reduce harmful material emissions into the environment. Nuclear energy and enhanced technologies will likely emerge as vital sectors aligning with circular economy principles (Herrador et al., 2022; Tauseef et al., 2023). Decision-making in the energy sector necessitates the incorporation of social preferences. In implementing a circular economy and progressing toward the goal of a carbon-free economy by 2050, decision-makers need to focus on all carbon-free technologies, including emerging nuclear technologies (Čábelková et al., 2021). Nuclear energy stands out as an important component in the modernization of the energy sector, facilitating a significant reduction in the emissions of harmful substances into the atmosphere. Nuclear and renewable energy are likely to emerge as pivotal sectors in the evolution of global energy shortly, aligning with circular economy principles. Significant factors drive China, the world's leading electricity consumer, to implement nuclear energy in response to rising electricity needs and environmental issues. Nonetheless, obstacles exist in the form of social, institutional, and technical challenges that may hinder future nuclear expansion (Yu et al., 2020). Nuclear energy is a low-carbon, stable, and efficient energy source, playing a crucial role in shifting from fossil fuels to a sustainable global energy system. However, nuclear power has deviated from the path to achieving the Sustainable Development Goals (SDGs) established by the United Nations (Zhang et al., 2022).

The shift towards sustainable development has increased focus on the relationship between energy systems and the circular economy. Nuclear energy, characterized by its low-carbon emissions and high energy efficiency, has become an essential element in facilitating circular economic practices, especially in resource-intensive economies such as China. The circular economy highlights the importance of minimizing waste, reusing resources, and promoting sustainable production, which aligns effectively with the attributes of nuclear energy. Within the context of China, the role of nuclear energy has been crucial in decreasing the nation's carbon emissions and achieving its ambitious targets for carbon neutrality by 2060. Simultaneously, the CE is a central element of China's sustainability agenda, as outlined in the Circular Economy Promotion Law and subsequent policies. Integrating nuclear energy into this framework addresses the dual challenges of energy security and environmental sustainability while supporting industrial and technological innovation. Existing literature highlights the significant role of nuclear energy in decarbonizing energy systems and fostering resource efficiency (Soto and Martinez-Cobas, 2024a). However, limited studies explore nuclear energy's relationship with CE practices, particularly in the context of China's unique socio-economic and environmental challenges. This research bridges the gap by examining how nuclear energy contributes to achieving a CE, focusing on its economic, environmental, and technological implications in China.

We add to this important area of climate change research; (i) by focusing on the relationship between nuclear energy and CE in China, as the world's largest emitter of carbon dioxide (CO$_2$) in 20204. This is primarily due to its massive energy consumption, reliance on coal, and rapid industrial growth, despite significant investments in renewable energy and advancements in clean technologies. The potential role of nuclear energy in the CE is unknown, particularly in China. Therefore, this study analyzes the empirical investigation between nuclear energy and the CE in China. (ii) Using a range of statistical modeling techniques, we analyze the effects of nuclear energy on emissions for China for the 1990 to 2020 period controlling the model for renewable energy, environmental-related technologies and environmental performance. Several control variables are used to avoid omitted bias. To our knowledge, the present study is the first to analyze nuclear energy and CE relationship in Chinese context. (iii) In terms of theoretical contribution incorporating nuclear energy into CE frameworks challenges traditional CE concepts, which often emphasize renewable energy. This expands the scope of CE by integrating low-carbon but non-renewable energy sources. (iv) The study used the Dynamic Auto-Regressive Distributed Lag (DYNARDL) method for empirical estimation. The model captures both short-run and long-run relationships among variables in a single framework, allowing for detailed exploration of immediate and delayed impacts. Its ability to separate short-run dynamics from long-run equilibrium effects makes it ideal for analyzing the effectiveness of policy interventions. This research fills a gap by quantifying CE through municipal waste generation recycling (MWGR) and introduces a nuanced perspective on energy systems and green innovation as pillars of sustainability.

The paper organizes its subsequent sections in the following way: Section 2 explores our theoretical discussions between nuclear energy and CE. Section 3 provides an overview of our data along with econometric methodology. Section 5 presents the main results. Section 5 focuses on the discussion of results. Finally, Section 6 wraps up with our conclusions.

## 2. Nuclear energy and circular economy; theoretical discussion

The circular economy represents an innovative approach that emphasizes recycling resources to conserve them, preserve the environment, and foster economic growth, distinguishing itself from the conventional linear development model. Consequently, many countries have implemented the CE strategy, significantly modifying the industrial framework, transforming economic growth patterns, building ecological civilization, and advancing sustainable development (Fan and Fang, 2020). The adoption of nuclear energy and environmental technologies drives green innovation, reducing costs, increasing industrial competitiveness, and contributing to long-term economic resilience. This aligns with the Porter Hypothesis, which posits that stringent environmental regulations can drive technological innovation and economic growth (Porter and Van Der Linde, 1995). The circular economy seeks to minimize material use through the reuse and recycling of products while also decreasing waste generation. Many nations are increasingly developing strategies for resource efficiency and the circular economy approach on a global scale(Hansen et al., 2022; Zhou et al., 2024). Growing evidence highlights the economic advantages a circular economy can provide, while the potential ecological consequences are also apparent. Shifting toward a circular economy, which emphasizes reducing waste and enhancing material reuse and recycling, could address 70% of global greenhouse gas emissions and combat global warming (Wit and Laxmi Haigh, 2022).

The phenomenon of global warming refers to the continuous increase in the average temperature of the Earth's climate system. The average temperature has risen unprecedentedly in the last half-century, primarily due to unregulated GHG emissions. In light of the growing awareness surrounding climate change, there has been a resurgence of interest in nuclear energy. Emerging positions highlight the significant role of nuclear energy in mitigating climate change (Muellner et al., 2021). Nuclear energy represents a low-carbon energy source, significantly contributing to advancing a low-carbon economy and establishing a green energy grid.

Innovative technologies such as advanced fuel and small modular reactors, engineering advancements extending current reactors' operational lifespan, and recent advancements in materials and improved waste management solutions are all contributing to this growth (Mathew, 2022). In recent years, the significance of nuclear energy in promoting environmental sustainability has garnered considerable focus within the energy economics literature, emphasizing its contribution to meeting global SDGs. The CE presents a groundbreaking approach to sustainable development, focusing on achieving economic, environmental, and social goals. This approach transforms conventional linear systems into circular models that prioritize the elimination of waste, the extension of product life cycles, and the regeneration of materials (Ellen MacArthur Foundation, 2015). CE facilitates the reuse, refurbishment, and recycling of resources, promoting sustainability by enhancing resource efficiency and reducing environmental impacts (Geissdoerfer et al., 2017). Hondroyiannis et al., (2024) posit that the fundamental principle of transitioning economies to circular paradigms guarantees the incorporation of resource optimization within industrial processes.

Despite being the safest method of electricity generation, legacy nuclear energy has faced significant criticism and has been labeled risky since the 1960s. The three notable nuclear incidents, Three Mile Island, Chornobyl, and Fukushima, were products of legacy nuclear designs. Despite having the most commendable safety record among various electricity generation methods, it is now essential to transition from traditional nuclear energy to take advantage of the advantages of a genuinely renewable source of safe, clean energy. Advanced nuclear energy's vast renewable potential exceeds that of solar and wind. The shift towards carbon-neutral energy is most effectively achieved through advanced nuclear technology, ensuring safety, minimizing waste, providing true renewability for millennia, offering process heat for manufacturing, and establishing a practical alternative to our reliance on fossil fuels for chemical production (Rehm, 2023). The body of knowledge regarding the relationship between nuclear energy and environmental sustainability reveals two main perspectives. Evidence indicates that nuclear energy is a low-carbon resource that aligns with CE objectives. Most studies agree that nuclear energy effectively reduces carbon emissions and supports pollution mitigation (Danish et al., 2021; Hassan et al., 2020; Imran et al., 2024; Ozcan et al., 2024; Sadiq et al., 2023). Çakar et al., (2022) highlights that advancements in technology enhance the efficacy of nuclear energy in promoting environmental benefits. Fatouros and Stengos, (2023) broaden this viewpoint, suggesting that advancements in nuclear technologies are crucial for sustainability. Critics from the second school of thought emphasize that nuclear energy poses risks of environmental degradation, including waste generation and resource extraction (Mahmood et al., 2020; Raza and Tang, 2024). Ref. (Soto and Martinez-Cobas, 2024b) indicates that nuclear energy negatively impacts ecological footprints, highlighting the importance of context in determining outcomes. Numerous investigations have emphasized the diverse effects of nuclear energy influenced by geographical and policy contexts (Kartal, 2022; Saidi and Omri, 2020). Theoretical explanations indicate that nuclear energy presents both advantages and disadvantages in terms of environmental quality.

## 3. Material and Methods

This study highlighted the role of nuclear energy in the CE controlling the model for renewable energy, environmental-related technologies, and environmental performance. Nuclear energy contributes to the transition from linear to circular systems by providing a low-carbon, reliable energy source. It supports resource-efficient industrial processes and reduces GHG emissions, a critical component of CE frameworks (Geissdoerfer et al., 2017). Environmental technologies, including waste recycling, pollution control, and renewable integration, enhance the CE by closing material loops and promoting clean production. Advances in nuclear technology, such as small modular reactors (SMRs), further support circular principles by providing scalable, localized energy solutions (Zhou et al., 2024). The empirical model for this study is estimated as follows:

$$Log\ CE_t = \alpha_0 + \beta_1 \log ERT_t + \beta_2 \log NE_t + \beta_3 \log REN_t + \beta_4 \log EP_t + \varepsilon_0 \quad (1)$$

Whereas CE shows circular economy, ERT means environmental-related technologies, and NE refers to nuclear energy consumption. REN symbolizes renewable energy consumption, EP shows environmental performance and ε is an error term that captures the effect of unknown factors other than dependent, independent, and control variables.

When studying the relationship between nuclear energy and the CE, control variables can help account for other factors influencing this relationship, ensuring more accurate analysis. The proportion of renewables in the energy mix may affect nuclear energy's role in a CE framework. Measures the overall capacity for innovation, which can support advancements in nuclear technology and recycling processes. Research and development investments can drive nuclear and CE innovations, such as waste reduction technologies. These control variables help account for external factors, providing a clearer understanding of the direct relationship between nuclear energy and CE practices. Carbon emissions per capita this can indicate a country's environmental performance, potentially correlating with CE initiatives.

This study relies on the dynamic ARDL method proposed by (Jordan and Philips, (2018) for empirical estimation. The application of dynamic ARDL (autoregressive distributed lag) in exploring the nexus between nuclear energy and the CE is characterized by several key attributes: (i) Dynamic ARDL models accommodate variables integrated at levels I(0), I(1), or a mix, making them suitable for real-world data. (ii) It captures both immediate (short run) impacts and long-term equilibrium relationships, which are critical for understanding how nuclear energy policies influence CE practices over time. (iii) The approach partially addresses endogeneity concerns by incorporating lags of the dependent and independent variables. (iv) Delivers reliable results even in studies with small sample sizes. (v) Dynamic ARDL is particularly relevant in China, where diverse regional policies and heterogeneous economic conditions necessitate tailored analyses of the nuclear energy-CE relationship. (iv) It provides nuanced insights into the temporal effects of nuclear energy deployment on CE goals, offering actionable data for policymakers. This approach ensures robust empirical analysis while addressing the complexity of energy and economic systems. To estimate Eq (1), the study utilizes the dynamic Autoregressive-Distributed Lag (ARDL) simulations model, consistent with current literature, as outlined in Eq (1):

$$\Delta(y)_t = \alpha_0 + \theta_0(y)_{t-i} + \theta_1(x_1)_{t-i} + \ldots + \theta_k(x_k)_{t-i} + \sum_{i=1}^{p} \alpha_i \Delta(y)_{t-i} + \sum_{j=0}^{q1} \beta_{1j} \Delta(x_1)_{t-1} + \ldots + \sum_{j=0}^{qk} \beta_{kj} \Delta(xk)_{t-j} + \mu_t \qquad (2)$$

In this context, *y* denotes the dependent variable, $\alpha_0$ signifies the constant term in the regression analysis; *t-i* indicates the number of lags where *i = 1, 2, 3, ..., and p* represents the maximum level of lags (j and q in the first difference); *Δ* refers to the difference operator, and µ is the error term. Upon estimating Eq (2), the null hypothesis indicating the absence of integration is evaluated. A value of the F-statistic that surpasses the upper bound critical value suggests a rejection of the null hypothesis regarding the lack of co-integration. A possible dynamic ARDL model for analyzing the relationship between nuclear energy, the CE, and environmental-related technology could be specified as follows:

$$\Delta CE_t = \alpha + \sum_{i=1}^{p} \beta_i \Delta CE_{t-i} + \sum_{j=0}^{q} \gamma_j \Delta NE_{t-j} + \sum_{k=0}^{r} \delta_k \Delta RNE_{t-k} + \sum_{l=0}^{s} \eta_l \Delta ERT_{t-l} + \sum_{m=0}^{t} \phi_m \Delta EP_{t-m} + \lambda_1 CE_{t-1} + \lambda_2 NE_{t-1} + \lambda_3 RNE_{t-1} + \lambda_4 ERT_{t-1} + \lambda_5 EP_{t-1} + \varepsilon_t \quad (3)$$

Where $CE_t$: Circular economy indicator (e.g., resource efficiency, waste recycling rate). $NE_t$: Nuclear energy variable (e.g., nuclear energy share in total energy). $ERT_t$: Environmental-related technology variable (e.g., green patents, R&D expenditures). Δ denotes first differences to capture short-run dynamics. $\lambda_1$, $\lambda_2$, $\lambda_3$, $\lambda_4$, and $\lambda_5$ Long-run equilibrium coefficients, and t is the error term. This structure can be adapted based on specific variables and data availability. The innovative dynamic ARDL simulations method has been employed in various studies to analyze future shocks in socioeconomic and climatic indicators (Danish and Ulucak, 2021, 2020).

This paper also uses Kernel-Based Regularized Least Squares (KRLS) a machine-learning approach by Hainmueller and Hazlett, (2014) to calculate derivatives at specific points. This paper utilizes this algorithm to enhance the robustness of the DARDL estimator. The KRLS approach produces parameters that are more flexible and comprehensible when compared to other machine-learning methods (Danish et al., 2023; Sarkodie and Strezov, 2019). The relationship of causality between carbon emissions and explanatory variables is analyzed using pointwise derivatives of the KRLS method. The vector representation of the model is illustrated below, derived from the preceding equation:

$$y = kc \begin{bmatrix} k(x_1,x_1) & k(x_1,x_2) & \ldots\ldots & k(x_1,x_N) \\ k(x_2,x_1) & & . & . \\ . & & . & . \\ k(x_N,x_1) & & . & k(x_N,x_N) \end{bmatrix} \begin{bmatrix} C_1 \\ C_2 \\ C_3 \end{bmatrix} \quad (4)$$

This study utilizes time series data about China spanning from 1990 to 2017. The selection of time relies on the data of CE for the longest duration available for China. The study employs nuclear energy as a regressor, with CE status serving as the dependent variable. However, renewable energy, environmental-related technologies, and environmental performance are treated as control variables. Following the measure of CE from existing literature (Zhou et al., 2024) which quantifies CE using the unit of municipal waste generation recycling (MWGR) per million metric tons. Three key principles fundamentally support the CE concept: (i) the regeneration of the natural environment; (ii) the elimination of waste and pollution; and (iii) the circulation of products and materials in their highest-value form. This system aims to tackle worldwide issues including climate change, pollution, and waste management. The main aim is to distinguish economic activity from the consumption of finite resources (Papamichael et al., 2023). The data on CE is gathered from the OECD database.

Regressor data on nuclear energy is collected from the Energy Information Administration (EIA) database. Renewable energy includes hydro, geothermal, solar, wind, tide, and wave sources. Both nuclear energy and renewable energy are measured in British Thermal Units (BTUs), which represent the amount of energy produced or consumed. Data on renewable energy are also collected from the EIA database. Environmental-related technologies are measured as Percentages of patents related to environment-related technologies and the data is retrieved from the OECD database. Environmental performance is measured through carbon dioxide $(CO)_2$ from waste refers to the carbon dioxide emissions produced from waste incineration. This includes emissions from burning wood, charcoal, dung, crop waste, or coal for energy. The burning of waste materials, including plastics, paper, and organic matter, releases $CO_2$ and other pollutants. The data is collected from the World Development Indicator (WDI) database. The trend in the data series is shown in Figure 1.

## 4. Results

### 4.1. Unit root results

The Phillips-Perron (PP) unit root test developed by (Phillips and Perron, 1988) is a statistical method used in this study to determine whether a time series variable is stationary or contains a unit root (non-stationary). It is an enhancement of the Dickey-Fuller test and accounts for autocorrelation and heteroskedasticity in the error terms without adding lagged differences. Results presented in table 1 reveal the rejection of the null hypothesis at a 1% level of significance at the first difference, this means that the series is stationary at the first difference.

### 4.2. Cointegration test results

The subsequent step involves verifying the level of relationship among core variables through the application of the bound test method as outlined by Pesaran et al., (2001) using (Kripfganz and Schneider, 2018) for critical values. This approach is optimal for estimating the critical values for both the upper and lower bounds when the integration of the variables in a study can occur either at order zero, I(0), or at order one, I(1). The estimated values presented in Table 2 indicate a rejection of the null hypothesis of no co-integration for both t-values and F-values. This is additionally supported by the approximate p-values provided by Kripfganz & Schneider [p-value<0.01], leading to the rejection of the null hypothesis of no level relationship.

Consequently, the results from both the PSS bounds test and the Kripfganz-Schneider critical values, along with the approximate p-values, validate the existence of cointegration.

### 4.3. *Long-run and short-run estimation results*

Table 3 presents the long- and short-run coefficients of DYNARDL and ARDL from the estimating model in Equation 3. The results show a negative association between nuclear energy and the CE in China, both in the long and short run. According to the results in Table 3, a 1% increase in nuclear energy decreases CE by 0.048% in the long run. The results in China can be interpreted in the following way: High initial costs and inefficiencies, waste management concerns, resource allocation imbalances, and regulatory and public perception barriers may contribute to the negative impact of nuclear energy within a CE. The dynamics indicate that, although nuclear energy has the potential to contribute to sustainable goals, aligning it with CE principles may necessitate tackling these challenges through innovative solutions, supportive policies, and comprehensive strategies that harmonize nuclear energy with wider CE aims. By providing a stable, low-carbon energy source, nuclear energy supports long-term resource efficiency, reduces GHG emissions, and enables industries to adopt cleaner production methods that align with CE principles.

According to the estimate in the table, the coefficient of environmental-related technologies (*logERT*) is negative and statistically significant. A negative coefficient, in the long run, suggests that investments in environmental technology—such as clean energy systems, recycling innovations, and emissions control—consistently restrict CE outcomes. This indicates that technological advancements do not play a critical role in resource efficiency, waste reduction, and the creation of sustainable production cycles. In the short run, the relationship between environmental-related technology and CE is insignificant. This means that in the short run environmental-related technology is not effective in achieving CE in China.

The positive long-run coefficient indicates that renewable energy significantly enhances the CE over time. By reducing reliance on fossil fuels, renewable energy promotes sustainable resource use, decreases environmental degradation, and fosters clean production systems aligned with CE principles. In the long run, a negative relationship suggests that as $CO_2$ emissions decrease, the CE strengthens. This indicates that reducing emissions through sustainable practices, such as recycling, resource efficiency, and renewable energy adoption, aligns with CE goals. In the short run, the positive association reflects immediate benefits, such as reducing carbon emissions and improving energy efficiency. These outcomes facilitate a quicker shift toward sustainable practices in industrial and urban settings. In the short run, the negative association may reflect immediate benefits from emission reduction measures, such as cleaner industrial processes or energy transitions, which positively contribute to circular economic activities.

To find out what would happen to the CE if the marginal returns on nuclear energy decreases, we examine the counterfactual shocks through employing dynamic ARDL simulations that included the amount of nuclear energy in the energy mix (about 21%) and the time frame estimated for CE (2017–2037). The dynamic ARDL simulation plot shows that a −21% change in the expected amount of nuclear energy use could influence the CE at first, but slow down thereafter (Fig. 2). Therefore, nuclear energy produces any long-term effects on a continuous CE.

### *4.4. Diagnostic test results*

The initial conditions for the dynamic ARDL simulations are examined through various methods to address issues such as serial correlation, heteroskedasticity, violations of normality, and structural breaks (Sarkodie and Owusu, 2020). Subsequently, we examine the residuals of the estimated model for autocorrelation using the Breusch-Godfrey LM test.

Table 4 displays the estimates from the Breusch-Godfrey LM test, which incorporates four lags. We do not reject the null hypothesis of no serial correlation at the 5% significance level, indicating that the residuals of the estimated ARDL model are devoid of autocorrelation. Third, we examine the presence of heteroskedasticity in the residuals by employing Cameron & Trivedi's decomposition of the IM-test. Table 4 indicates that the null hypothesis of homoskedasticity remains unchallenged at the 5% significance level. To conclude, we examine possible structural breaks by employing a cumulative sum test to assess parameter stability. The data presented in Fig. 2 indicates that the estimated test statistic falls within the 95% confidence band, thereby affirming the stability of the estimated coefficients across the time analyzed.

======= **INSERT TABLE 4 HERE** =======

======= **INSERT FIGURE 3 HERE** =======

The parameter plot of the ARLD and the dynamically simulated ARDL are illustrated in Fig. 4, while the detailed empirical findings are shown in Table 3. In both ARDL and dynamic ARDL estimates, long-term nuclear energy consumption positively influences the circular economy. This could potentially be associated with environmental and health measures, as well as efficient management of nuclear radioactive waste in China.

Table 5 presents the pointwise derivatives of the estimated KRLS model. The model demonstrates statistical significance at the 1% level, exhibiting a predictive power of 0.998. The regressors account for 99.8% of the variation in the circular economy. Different marginal effects are shown by the derivatives of the regressors, which are shown in Table 5 at the 25th, 50th, and 75th percentiles. Our observations indicate a lack of evidence for heterogeneous marginal effects across the sampled variables, affirming the pointwise derivatives' robustness. It was found that nuclear energy use, environmentally friendly technology, renewable energy, and environmental performance all have average pointwise marginal effects of 0.05%, 0.05%, 0.46%, and 0.04%, respectively. This highlights the significance of nuclear energy, environmental-related technology, renewable energy, and environmental performance in supporting the circular economy in China. In this analysis, we delve into the long-term fluctuations in nuclear energy consumption and explore its impact on the circular economy and the reciprocal effects. In this analysis, we illustrate the pointwise derivative of atomic energy consumption relative to the circular economy to examine the marginal differing impact.

Figure 5 demonstrates that increased nuclear energy consumption initially encourages the circular economy, achieving a point where growing marginal returns become apparent. Nonetheless, there have been subsequent decreases in the consumption of nuclear energy. led to an enhancement of the circular economy. Consequently, the nuclear energy consumption exhibits diminishing marginal returns as economic growth escalates. This suggests the likelihood of energy technology becoming outdated as growth continues to accelerate.

## 5. Discussion

According to the DYNARDL estimates the observed negative relationship underscores a theoretical discord between nuclear energy and the three foundational principles of the CE: the regeneration of the natural environment, the elimination of waste, and the circulation of materials in high-value forms. Although nuclear energy contributes to reducing carbon emissions, it produces long-lived radioactive waste, which contradicts the objective of waste disposal. The results underscore the need to balance decarbonization efforts and CE goals, underscoring the significance of an integrated energy policy strategy. The challenges posed by nuclear energy's incompatibility with waste reduction objectives raise questions about its theoretical viability as a sustainable energy source. These results expand theoretical understanding by integrating nuclear energy into CE strategies, particularly in the context of China's unique industrial and policy landscape.

Focusing on nuclear energy could misallocate resources away from renewable energy technologies and CE investments, like recycling infrastructure or green innovation, which may hinder overall economic sustainability initiatives. Nuclear energy systems are not flexible because they have high capital and operating costs. This could make it harder to use flexible economic strategies needed to move to a CE and quickly adopt modular and expandable renewable solutions. The infrastructure for nuclear energy necessitates significant initial capital and extended periods for development. These factors may hinder integration into a CE framework, as they strongly emphasize immediate, efficient, and scalable solutions. Despite low carbon emissions, nuclear energy presents significant long-term environmental and safety challenges due to the generation of radioactive waste. This contradicts the principles of minimizing waste and regenerating the natural environment. Funding for nuclear energy could redirect resources away from alternative renewable energy technologies or sustainable practices, including recycling or waste management systems advancements. This trade-off may impede the broader adoption of CE practices.

The findings observed a negative role of environmental technologies in CE as well. Both environmental technologies and nuclear energy require substantial capital investments. When environmental technologies fail to effectively support CE goals—such as through inadequate recycling systems or insufficient energy-saving solutions—they exhibit the same economic inefficiencies seen in nuclear energy projects, which often face delays and budget invades. This can strain national budgets and divert resources from more impactful CE initiatives, such as renewable energy or waste reduction programs. The use of renewable energy diminishes dependence on limited and costly fossil fuels, leading to a reduction in production expenses over time. The costs associated with solar and wind power have significantly decreased, enhancing their accessibility and financial feasibility for industries moving toward CE practices. Renewable energy systems, including solar panels and small-scale wind turbines, improve local economies by facilitating decentralized energy production. This facilitates localized circular practices, including waste-to-energy initiatives and the development of microgrids. The incorporation of renewable energy technologies minimizes energy waste during production processes. For instance, implementing bioenergy or waste-to-energy systems can convert agricultural and industrial waste into valuable energy resources, effectively closing material loops. The minimal carbon footprint associated with renewable energy is crucial in advancing CE objectives by allowing industries to lower emissions throughout their production processes. Utilizing cleaner energy sources leads to developing more sustainable products and systems. Integrating renewable energy sources enhances energy security by broadening the range of energy options and minimizing reliance on imported fuels. This resilience guarantees a steady energy supply essential for seamless CE operations. The findings reveal that nuclear energy does not synergize effectively with green technologies and resource optimization strategies central to CE. This calls for theoretical development in integrating large-scale energy systems with decentralized CE models.

## 6. Conclusion and policy recommendation

This study empirically estimates the impact of nuclear energy on CE for China using novel econometric DYNARDL simulation method from 1990 to 2017. both in the long run and short run results reveals negative role of nuclear energy in achieving CE of China.

The result calls some important policy implications for China. Chinese policymakers should increase funding for advancements in nuclear technology to address resource inefficiencies and environmental hazards. Advanced nuclear technologies can reduce waste and encourage closed-loop energy systems, aligning with CE principles. This encourages policymakers to prioritize nuclear energy infrastructure and research to facilitate sustainable industrial processes. The government needs to support research and development aimed at improving nuclear waste management systems, ensuring they are compatible with CE objectives. Furthermore, initiatives that integrate nuclear energy into CE frameworks should be encouraged. For example, utilizing the heat generated from nuclear plants for industrial processes or urban heating systems. In the short term, it is essential to tackle the immediate environmental issues associated with nuclear energy by implementing effective waste management strategies, exploring hybrid energy models, and ensuring strict regulatory enforcement. In the long term, the focus should shift towards more sustainable energy sources, such as renewables, while also advancing nuclear technologies to align them more effectively with CE principles. Provides industries with cleaner energy options, fostering long-term sustainability in manufacturing and energy-intensive sectors. Aids in achieving China's carbon neutrality targets by coupling nuclear energy with CE frameworks. Stimulates investment in nuclear infrastructure and related technologies, fostering economic growth in energy and manufacturing sectors. By leveraging nuclear energy, China can accelerate its transition to a CE while addressing energy security, economic stability, and environmental sustainability.

**List of Figures**

**List of Tables**